\UseRawInputEncoding
\documentclass[prd,aps,preprint,amsmath,nofootinbib,amssymb,eqsecnum,showkeys,tightenlines]{revtex4-1}

\usepackage{amsmath, latexsym, amssymb, hyperref, graphicx, color, multirow, makecell, diagbox}
\usepackage[table]{xcolor}
\usepackage{tabularx}
\usepackage[T1]{fontenc}
\usepackage[capitalize]{cleveref}
\definecolor{nicered}{rgb}{.7,.1,.1}
\definecolor{nicegreen}{rgb}{.1,.5,.1}
\definecolor{darkblue}{rgb}{0,0,.5}
\hypersetup{colorlinks, citecolor=nicegreen,linkcolor=nicered, urlcolor=darkblue}
\usepackage{multirow}
\usepackage{slashed}
\numberwithin{equation}{section}
\usepackage{verbatim}
\allowdisplaybreaks

\begin{document}
\preprint{}

\title{Radiative Leptonic Decay of Heavy Quarkonia}

\author{Junle Pei$^{1,2}$}
\email{peijunle@ihep.ac.cn}
\thanks{Corresponding author.}

\author{Xinchou Lou$^{1,3,4}$}
\email{xinchou@ihep.ac.cn}

\author{Yaquan Fang$^{1,3}$}
\email{fangyq@ihep.ac.cn}

\author{Jinfei Wu$^{1,5}$}
\email{wujf@ihep.ac.cn}

\author{Manqi Ruan$^{1,3}$}
\email{ruanmq@ihep.ac.cn}

\affiliation{$^1$ Institute of High Energy Physics, Chinese Academy of Sciences, Beijing 100049, People's Republic of China}
\affiliation{$^2$ Spallation Neutron Source Science Center, Dongguan 523803, People's Republic of China}
\affiliation{$^3$ University of Chinese Academy of Sciences, Beijing 100049, People's Republic of China}
\affiliation{$^4$ University of Texas at Dallas, Richardson, TX, USA}
\affiliation{$^5$ China Center of Advanced Science and Technology, Beijing 100190, People's Republic of China}

\begin{abstract}
This study examines the properties of heavy quarkonia $X$ by treating them as bound states of $Q$ and $\bar{Q}$ at the LO level within the NRQCD framework, where $Q$ represents either a charm or a bottom quark. The branching ratios for the radiative leptonic decays $X\rightarrow \gamma l^{+} l^{-}$ are revisited and the angular and energy/momentum distributions of the final state particles are analyzed in the rest frame of $X$. Furthermore, we apply Lorentz transformations from the rest frame of $X$ to the center-of-mass frame of $l^+ l^-$ to establish the connection between the widths ${\Gamma_{X \rightarrow \gamma l^{+} l^{-}}}$ and ${\Gamma_{X \rightarrow l^{+} l^{-}}}$. When comparing the connection with those documented in the literature (divided by $2\pi$) for various $X$ states, such as $J/\Psi$, $\Psi(2S)$, $\Upsilon(1S)$, and $\Upsilon(2S)$, relative differences typically around or below 10\% can be found, which is comparable to the NLO corrections of $O(\alpha)$ and $O(v^4)$. However, we observe a significant disparity in the ratio between ${\Gamma_{\Psi(2S) \to \gamma \tau^+ \tau^-}}$ and ${\Gamma_{\Psi(2S) \to \tau^+ \tau^-}}$, with our prediction being four times larger than those in the literature. The outcomes derived from this study held practical implications in describing the QED radiative processes and contribute to the investigation of QCD processes associated with the decays of heavy quarkonia  and the searches for new physics.
\end{abstract}

\maketitle

\section{Introduction}\label{Section1}
	
	The radiative leptonic decays of heavy quarkonia, denoted as $X$, specifically $X\to \gamma l^+ l^-$ where $l=e, \mu, \tau$, play a crucial role in tests of Quantum ElectroDynamics (QED) due to the absence of hadronic final states. Moreover, in studies of Quantum ChromoDynamics (QCD) and searches for physics beyond the Standard Model (SM) at colliders, where copious hadronic events are produced, accurate simulation of QED radiative background events is essential. This necessitates the utilization of software tools, such as PHOTOS \cite{Barberio:1990ms,Barberio:1994qi,Dobbs:2004qw}, which rely on appropriate theoretical inputs regarding  not only the decay branching ratios of $X$ but also the differential distributions of the final state particles. However, only the branching ratio for the decay $J/\Psi\to \gamma e^+ e^-$ is available in the Particle Data Group (PDG) \cite{ParticleDataGroup:2022pth} when considering $X$ states of quantum configuration $I^G(J^{PC})=0^- (1^{--})$, such as $J/\Psi$, $\Psi(2S)$, $\Upsilon(1S)$, and $\Upsilon(2S)$. This value was obtained based on the predicted ratio between ${\Gamma_{J/\Psi \rightarrow \gamma e^{+} e^{-}}}$ and ${\Gamma_{J/\Psi \rightarrow e^{+} e^{-}}}$ \cite{FermilabE760:1996jsx}. Consequently, it is imperative to encourage further experimental measurements and accurate theoretical predictions for the radiative leptonic decays of heavy quarkonia.
	
	In Ref. \cite{FermilabE760:1996jsx}, the ratio between ${\Gamma_{X \rightarrow \gamma l^{+} l^{-}}}$ and ${\Gamma_{X \rightarrow l^{+} l^{-}}}$ was calculated  based on the QED formalism presented in Ref. \cite{book:JM}. The ratio is defined using quantities in the center-of-mass (c.m.) frame of the $l^+ l^-$ system. By utilizing the measured value of ${\Gamma_{X \rightarrow l^{+} l^{-}}}$ and the aforementioned ratio, it becomes possible to determine the radiative leptonic decay width ${\Gamma_{X \rightarrow \gamma l^{+} l^{-}}}$ and branching ratio without considering the wave functions of the bound state $X$ at the origin  explicitly. This ratio and results from Ref. \cite{FermilabE760:1996jsx} have also been employed in subsequent studies \cite{Li:2009wz,Rashed:2014kla,Huber:2012hsa,McGlinchey:2012lka} to investigate the radiative leptonic decays of quarkonia. However, it is important to note that the c.m. frame of the $l^+ l^-$ system varies with the energy and direction of the final state photon. As a result, the expression proposed in Ref. \cite{FermilabE760:1996jsx} does not directly provide information regarding the angular and energy/momentum distributions of the final state particles in the rest frame of $X$.
	
	This study focuses on the reexamination of the radiative leptonic decays of heavy quarkonia within the framework of Non-Relativistic QCD (NRQCD) \cite{Bodwin:1994jh}. For simplicity and to facilitate comparison with the results in the literature and PDG, this study does not include the processes of $X\to X^\prime \gamma \to \gamma l^+ l^-$, where $X^\prime$ is another resonance lighter than $X$. The heavy quarkonia $X$ is regarded as a bound state of a quark ($Q=c,b$) and its antiquark ($\bar{Q}$), and its wave function is analyzed in both momentum and position spaces. At the leading-order (LO) level, the decay width of the quarkonia $X$ into specific final states is proportional to the square of the wave function at the origin, denoted as $|\psi({\vec{0}})|^2$. As the ratio between different decay widths of $X$ decay modes is independent of $|\psi({\vec{0}})|^2$, we provide preditions for ${\Gamma_{X \rightarrow \gamma l^{+} l^{-}}}$ in both the rest frame of $X$ and the c.m. frame of $l^+ l^-$ by utilizing ${\Gamma_{X \rightarrow l^{+} l^{-}}}$ and the ratio ${\Gamma_{X \rightarrow \gamma l^{+} l^{-}}}/{\Gamma_{X \rightarrow l^{+} l^{-}}}$. Numerical integration in the rest frame of $X$ provides insight into the angular and energy/momentum distributions of the final state particles. On the other hand, the relation in the c.m. frame of $l^+ l^-$ yields a ratio between ${\Gamma_{X \rightarrow \gamma l^{+} l^{-}}}$ and ${\Gamma_{X \rightarrow l^{+} l^{-}}}$, which can be compared with the corresponding ratio proposed in Ref. \cite{FermilabE760:1996jsx}. The results can be incorporated into some generators, similar as PHOTOS, as theoretical inputs for simulating QED radiative processes during tests of the SM  and the search for new physics.

	The structure of this paper is organized as follows. In Section~\ref{Section2f}, a brief introduction to the production and decay of heavy quarkonia within the framework of NRQCD is provided. Section~\ref{Section2} presents an analytical description of the kinematics of the final state particles in the decay $X \to \gamma l^{+} l^{-}$, specifically focusing on the $X$ rest frame. In Section~\ref{Section3}, to illustrate the analysis, the decay $J/\Psi \to \gamma \mu^+ \mu^-$ is used as an example, and the angular and energy/momentum distributions of the final state particles are derived at the LO level within the NRQCD framework. Moving on to Section~\ref{Section4}, the Lorentz transformation from the $X$ rest frame to the $l^+ l^-$ c.m. frame is performed. This section presents the derivation of the ratio between ${\Gamma_{X \rightarrow \gamma l^{+} l^{-}}}$ and ${\Gamma_{X \rightarrow l^{+} l^{-}}}$, and a comparison is made between this proposed expression and the one presented in Ref. \cite{FermilabE760:1996jsx}. Finally, the paper concludes in Section~\ref{Section6}.
	
	\section{Heavy quarkonia production and decay within the framework of NRQCD}\label{Section2f}
	
	In this study, we use the symbol $X$ to represent heavy quarkonia of $Q\bar{Q}$ ($Q=c,b$) state. Within the framework of NRQCD \cite{Bodwin:1994jh}, the physics of $X$ production can be divided into a perturbative part and a non-perturbative part based on the energy scale.
	The perturbative part involves the production of a heavy-quark pair $Q\bar{Q}[n]$ with quantum configuration $n$ at a hard scale $\Lambda\sim m_Q$, where $n=~^{2S+1}L_J^{[1,8]}$ ($S$, $L$, and $J$ represent the total spin, the orbital angular momentum, and the total angular momentum of the $Q\bar{Q}$ pair respectively, and the $Q\bar{Q}$ pair is in either a color-singlet or color-octet state indicated by the superscript [1, 8]).
	The non-perturbative part refers to the hadronization of the $Q\bar{Q}[n]$ pair into the bound state $X$ at a significantly lower scale $\Lambda^\prime<m_Q$. 
This factorization can be expressed as \cite{Abreu:2022cco}	
    \begin{align}
	    d\hat{\sigma}_{ab}(X+\{k\})=\sum_n d\hat{\sigma}_{ab}(Q\bar{Q}[n]+\{k\}) \langle \mathcal{O}_X^n\rangle~, 
	\end{align}
where $d\hat{\sigma}_{ab}(Q\bar{Q}[n]+\{k\})$ describes the production of the heavy-quark pair $Q\bar{Q}$ of quantum configuration $n$ with additional final state partons $\{k\}$. $\langle \mathcal{O}_X^n\rangle$ represents the non-perturbative Long-Distance Matrix Element (LDME) that describes the hadronization of the $Q\bar{Q}[n]$ state into the bound state $X$.

Similarly, within the NRQCD framework, the decay of $X$ into light hadrons can also be factorized into two parts as \cite{Bodwin:1994jh}
\begin{align}
    \Gamma(X\rightarrow \text {LH})&=2 \text{Im}\langle X |\delta\mathcal{L}_\text{4-fermion}|X \rangle \nonumber\\
    &=\sum_n \frac{2\text{Im}f_n(\Lambda)}{m_X^{d_n-4}}
    \langle X |\mathcal{O}_n (\Lambda)|X \rangle~,\label{xxxp}
\end{align}
where $\delta\mathcal{L}_\text{4-fermion}$ represents the 4-fermion correction terms in the NRQCD Lagrangian. This includes the summation of different local 4-fermion operators $\mathcal{O}_n$ of quantum configuration $n$ multiplied by corresponding coefficients $f_n$. 
$d_n$ represents the naive scaling dimension of the operator $\mathcal{O}_n$. The coefficient $\text{Im}f_n$, which can be computed perturbatively, is proportional to the rates for on-shell $Q$ and $\bar{Q}$ to annihilate from the initial quantum configuration $n$ into the final states. The non-perturbative matrix elements $\langle X |\mathcal{O}_n|X \rangle$ indicate the probability of finding the $Q$ and $\bar{Q}$ in the quantum configuration $n$ within the bound state $X$, which is suitable for annihilation.

\begin{figure}[ht]
		\centering
		\includegraphics[width=.45\textwidth]{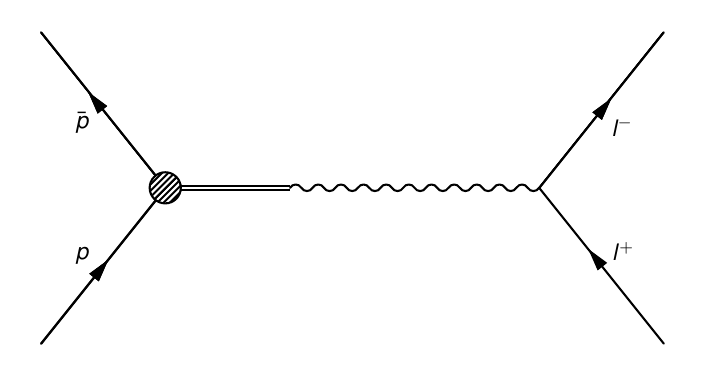}
		\caption{Diagrams for decay $X\longrightarrow  l^{+} l^{-}$.
		}
		\label{feynd2}
	\end{figure}

In this study, we only consider the (radiative) leptonic decay of $X$ with the quantum configuration $I^G(J^{PC})=0^- (1^{--})$, denoted as $X\longrightarrow (\gamma) l^{+} l^{-}$.
Within the NRQCD framework, the decay width can be expressed as \cite{Bodwin:1994jh}
\begin{align}
    &\Gamma(X\longrightarrow (\gamma) l^{+} l^{-})=
    \frac{2\text{Im}f_{(\gamma)l^+ l^-}(^3 S_1)}{m_X^2} |\langle 0|\chi^\dagger\vec{\sigma} \psi| \psi\rangle |^2  \nonumber \\
    & +\frac{2\text{Im}g_{(\gamma)l^+ l^-}(^3 S_1)}{m_X^4} \text{Re}\left( \langle \psi|\psi^\dagger\vec{\sigma} \chi|0\rangle\cdot 
    \langle 0|\chi^\dagger\vec{\sigma}\left(-\frac{i}{2}\mathop{\text{D}} \limits ^{\leftrightarrow}\right)^2 \psi|\psi\rangle\right) +O(v^4\Gamma)~,\label{nrqcd}
\end{align}
where $v$ represents the velocity of $Q$ ($\bar{Q}$) in the bound state $X$. The average value of $v^2$ is approximately 0.3 for charmonium and around 0.1 for bottomonium \cite{Quigg:1979vr}. The second term in Eq.~(\ref{nrqcd}) is proportinal to $v^2$. Since the processes of $X\longrightarrow l^{+} l^{-}$ and $X\longrightarrow (\gamma) l^{+} l^{-}$ share almost the same Feynman diagrams by using the optical theorem, as given in Ref.~\cite{Bodwin:1994jh}, we have
\begin{align}
    &\frac{\text{Im}g_{(\gamma)l^+ l^-}}{\text{Im}f_{(\gamma)l^+ l^-}}=-\frac{4}{3}+O(\alpha_S)~,\\
    &\frac{\text{Im}f_{\gamma l^+ l^-}}{\text{Im}f_{l^+ l^-}}=\frac{\text{Im}g_{\gamma l^+ l^-}}{\text{Im}g_{l^+ l^-}}\left(1+O(\alpha)\right)~,\\
\end{align}
where $\alpha$ and $\alpha_S$ are coupling constants of electromagnetic interaction and strong interaction, respectively. So, we get
\begin{align}
    \frac{\Gamma(X\longrightarrow \gamma l^{+} l^{-})}{\Gamma(X\longrightarrow l^{+} l^{-})}&=\frac{\text{Im}f_{\gamma l^+ l^-}}{\text{Im}f_{l^+ l^-}}\left(1+O(\alpha)+O(v^4)\right)\nonumber\\
    &=\frac{\text{Im}f^{\text{LO}}_{\gamma l^+ l^-}}{\text{Im}f^{\text{LO}}_{l^+ l^-}}\left(1+O(\alpha)\right)\left(1+O(\alpha)+O(v^4)\right)\nonumber\\
    &=\frac{\text{Im}f^{\text{LO}}_{\gamma l^+ l^-}}{\text{Im}f^{\text{LO}}_{l^+ l^-}}\left(1+O(\alpha)+O(v^4)\right)~,\label{nlo}
\end{align}
where $\text{Im}f^{\text{LO}}_{l^+ l^-}$ and $\text{Im}f^{\text{LO}}_{\gamma l^+ l^-}$ are calculated from diagrams in FIG.~\ref{feynd2} and FIG.~\ref{feynd}, respectively. 
Eq.~(\ref{nlo}) demonstrates that by only considering the diagrams in FIG.~\ref{feynd} and FIG.~\ref{feynd2} to calculate the ratio between $\Gamma(X\longrightarrow \gamma l^{+} l^{-})$ and $\Gamma(X\longrightarrow  l^{+} l^{-})$, the next-to-leading order (NLO) corrections are of $O(\alpha)$ and $O(v^4)$.

\section{Analytical description of final state kinematics
	}\label{Section2}
	
	 For the radiative leptonic decay of a quarkonium, denoted as $X\longrightarrow \gamma l^{+} l^{-}$, the four-momenta of the final state particles in the rest frame of $X$ can be parameterized as follows:
	\begin{align}
	& p_{l^-}^\mu = (\sqrt{p^2_{l^-}+m_l^2},~p_{l^-}\sin\theta_{l^-},~0,~p_{l^-}\cos\theta_{l^-})~,\\
	& p_{\gamma}^\mu = (E_\gamma,~E_\gamma\sin\theta_\gamma\cos\phi_\gamma,~E_\gamma\sin\theta_\gamma\sin\phi_\gamma,~E_\gamma\cos\theta_\gamma)~,\\
	& p_{l^+}^\mu =p_X^\mu-p_{l^-}^\mu-p_{\gamma}^\mu~,\label{4ll}
	\end{align}
	where $p_X^\mu=(m_X,~0,~0,~0)$. Utilizing the on-shell condition of $l^+$, we obtain the relation
	\begin{align}
	\cos\phi_\gamma=\frac{m_X^2-2m_X E_\gamma -2(m_X-E_\gamma)\sqrt{p^2_{l^-}+m_l^2}}{2 p_{l^-} E_\gamma \sin\theta_{l^-} \sin\theta_\gamma}-\frac{1}{\tan\theta_{l^-} \tan\theta_\gamma}~.\label{phi}
	\end{align}
	The variables $p_{l^-}$, $E_\gamma$, $\theta_{l^-}$, and $\theta_\gamma$ are complete to describe the kinematics of the final state particles. 
	The sensible regions of the parameter space should ensure that $E_{l^+}$ obtained from Eq. (\ref{4ll}) is greater than $m_l$, and $\cos\phi_\gamma$ given by Eq. (\ref{phi}) ranges from $-1$ to $1$. Consequently, the accessible ranges of the parameters are determined as follows:
	\begin{align}
	& p_{l^-}\in (0,~p_{l\text{max}})~,\\
	& E_{\gamma} \in \begin{cases}\left(m_X\frac{\frac{m_X}{2}-\sqrt{p^2_{l^-}+m_l^2} }{m_X-\sqrt{p^2_{l^-}+m_l^2}+p_{l^-}},~ m_X\frac{\frac{m_X}{2}-\sqrt{p^2_{l^-}+m_l^2} }{m_X-\sqrt{p^2_{l^-}+m_l^2}-p_{l^-}}\right) & p_{l^-} \in\left(0, ~p_{l\text {mid }}\right) \\ 
	\left(E_{\gamma\text {cut }},~ m_X\frac{\frac{m_X}{2}-\sqrt{p^2_{l^-}+m_l^2} }{m_X-\sqrt{p^2_{l^-}+m_l^2}-p_{l^-}}\right) & p_{l^-} \in\left(p_{l\text {mid }},~ p_{l\text {max }}\right)\end{cases},\label{errr}\\
	&\theta_+ \in\left(\theta^{\prime}[p_{l^-}, E_{\gamma}], ~2 \pi-\theta^{\prime}[p_{l^-}, E_{\gamma}]\right)~,\\
	&\theta_{-} \in\left(-\theta^{\prime}[p_{l^-}, E_{\gamma}],~ \theta^{\prime}[p_{l^-}, E_{\gamma}]\right)~,
	\end{align}
	where $E_{\gamma\text{cut}}$ represents the photon energy cut (only considering photons with energy greater than $E_{\gamma\text{cut}}$), and
	\begin{align}
	&
	\begin{cases}
	p_{l\text{max} }=\frac{E_{\gamma\text{cut}}}{2}+\frac{m_X-E_{\gamma\text{cut}}}{2} \sqrt{1-\frac{4 m_l^2}{m_X^2-2 E_{\gamma\text{cut}} m_X}} \\
	p_{l\text{mid} }=-\frac{E_{\gamma\text{cut}}}{2}+\frac{m_X-E_{\gamma\text{cut}}}{2} \sqrt{1-\frac{4 m_l^2}{m_X^2-2 E_{\gamma\text{cut}} m_X}}
	\end{cases},\label{prange}\\
	&
	\begin{cases}
	\theta_{+}=\theta_{l^-}+\theta_\gamma\\
	\theta_{-}=\theta_{l^-}-\theta_\gamma
	\end{cases},\\
	&\theta^{\prime}[p_{l^-}, E_{\gamma}]=\cos ^{-1}\left(\frac{\frac{m_X^2}{2}-m_X E_{\gamma} -(m_X-E_{\gamma})\sqrt{p^2_{l^-}+m_l^2}}{p_{l^-} E_{\gamma}}\right)~.
	\end{align}
	
	\section{Angular and energy distributions of final states in the quarkonia rest frame}\label{Section3}
	
	At the LO level ($v=0$) within the NRQCD, we can consider $X$ with the quantum configuration $I^G(J^{PC})=0^- (1^{--})$ as a bound state of $Q \bar{Q}[n]$ with $n=~^{3}S_1^{[1]}$. The momentum-space quantum state of $X$ can be expressed as \cite{Peskin:1995ev}
	\begin{align}
	& |X\rangle=\sqrt{2 m_X} \int \frac{d^3 \vec{k}}{(2 \pi)^3} \widetilde{\psi}(\vec{k}) \frac{1}{\sqrt{2 m_Q}} \frac{1}{\sqrt{2 m_Q}}|\vec{k},-\vec{k}\rangle~, 
	\end{align}
	where $\widetilde{\psi}(\vec{k})$ represents the wave function in momentum space, and $\vec{k}$ ($-\vec{k}$) is the three-momentum of $Q$ ($\bar{Q}$) in the rest frame of $X$, satisfying the condition $|\vec{k}|\ll m_Q$.
	The amplitude for $X$ to decay into specific final states is given by \cite{Peskin:1995ev}
	\begin{align}
	\mathcal{M}(X \rightarrow \text { final states }) & \approx \sqrt{\frac{2}{m_X}} \psi(\vec{0}) \mathcal{M}(\vec{0}, \vec{0} \rightarrow \text { final states })~. \label{m3}
	\end{align}
	To obtain Eq. (\ref{m3}), we approximately treat $Q$ and $\bar{Q}$ as static since $v=0$ ($|\vec{k}|=0$) at the LO level within the NRQCD. And we also use the approximation $2m_Q\approx m_X$. 
	In Eq. (\ref{m3}), $\psi(\vec{0})$ represents the wave function of $X$ at the origin ($\vec{0}$), obtained from the Fourier transformation of $\widetilde{\psi}(\vec{k})$ as
	\begin{align}
	\int \frac{d^3 \vec{k}}{(2 \pi)^3} \widetilde{\psi}(\vec{k})=\psi(\vec{0})~.
	\end{align}
	
	Consequently, the decay width of $X$ into specific final states can be approximated as \cite{Peskin:1995ev}
	\begin{align}
	\Gamma(X \rightarrow \text { final states }) 
	& \approx \frac{|\psi(\vec{0})|^2}{m_X^2} \int d \Pi_f \left|\mathcal{M}(\vec{0}, \vec{0} \rightarrow \text { final states })\right|^2~,\label{decayx}
	\end{align}
	where $\Pi_f$ represents the phase space of the final state particles.
{The correlation between Eq. (\ref{decayx}) and Eq. (\ref{xxxp}) is established through 
	\begin{align}
    & \langle X |\mathcal{O}_{^{3}S_1^{[1]}} (\Lambda)|X \rangle=|\psi(\vec{0})|^2~,
     \\
     & 2\text{Im}f_{^{3}S_1^{[1]}}(\Lambda)= \int d \Pi_f \left|\mathcal{M}(\vec{0}, \vec{0} \rightarrow \text { final states })\right|^2 ~.
\end{align}}
	Considering the parameters we have described, $d\Pi_f$ is given by
	\begin{align}
	d \Pi_f=\frac{1}{8(2\pi)^4}\frac{1}{\sqrt{1+\frac{m_l^2}{p^2_{l^-}}}}\frac{1}{\sqrt{1-\cos^2\phi_\gamma}}dp_{l^-}~dE_\gamma~d\theta_+ ~d\theta_{-}~,\label{ccc}
	\end{align}
	In Eq. (\ref{ccc}), $\cos\phi_\gamma$ is given by Eq. (\ref{phi}), with $\theta_l$ and $\theta_\gamma$ replaced by $\frac{\theta_{+}+\theta_{-}}{2}$ and $\frac{\theta_{+}-\theta_{-}}{2}$, respectively.
	
	\begin{figure}[ht]
		\centering
		\includegraphics[width=1.0\textwidth]{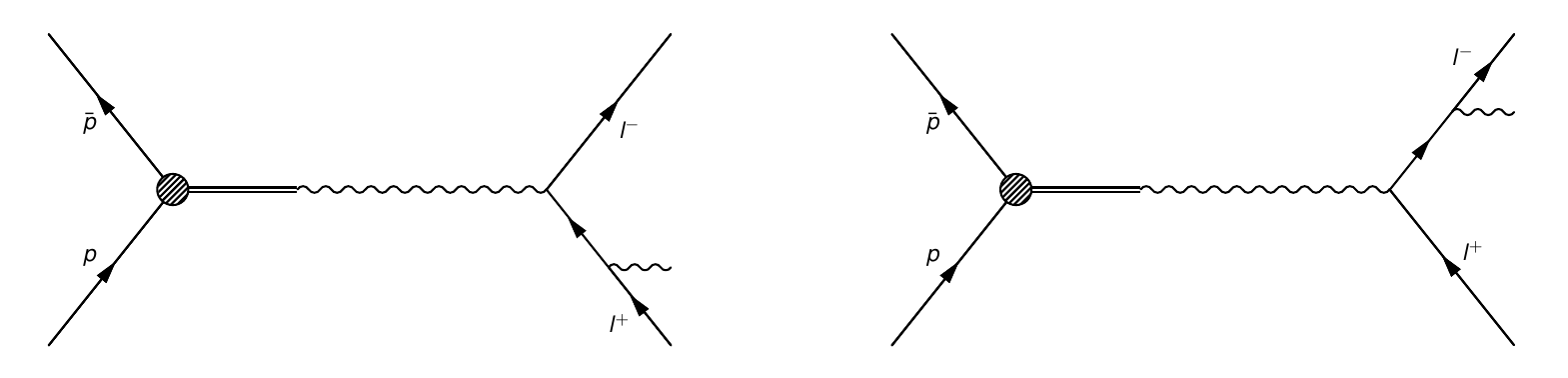}
		\caption{Diagrams for decay $X\longrightarrow \gamma l^{+} l^{-}$.
		}
		\label{feynd}
	\end{figure}

	In FIG.~\ref{feynd}, we present the diagrams for the decay $X\longrightarrow \gamma l^{+} l^{-}$. Since $m_X\ll m_Z, m_H$ (where $m_Z$ and $m_H$ represent the masses of the Z boson and the Higgs boson, respectively), we only consider the Feynman diagrams where the photon serves as the intermediate line for the decay $X\longrightarrow \gamma l^{+} l^{-}$. Additionally, due to charge-conjugation invariance \cite{FermilabE760:1996jsx}, the photon in the final state can only be emitted by one of the charged leptons in the final state. Therefore, we consider the diagrams shown in FIG.~\ref{feynd}. Direct calculation yields the following result:
	{\small 
		\begin{align}
		&\sum_{\lambda_Q,\lambda_{\bar{Q}}=\pm 1}\left|\mathcal{M}(\vec{0}, \vec{0} \rightarrow \gamma l^+ l^-)\right|^2=-\frac{4096 \pi^3 \mathcal{Q}_Q^2  \alpha^3}{m_X^4\left(m_X-2 \sqrt{p^2_{l^-}+m_l^2}\right)^2\left(m_X-2\left(\sqrt{p^2_{l^-}+m_l^2}+{E_\gamma}\right)\right)^2}\times \nonumber\\
		&\left(  2 p^2_{l^-} m_X\left(m_X\left(-8 m_X \sqrt{p^2_{l^-}+m_l^2}+8 {E_\gamma} \sqrt{p^2_{l^-}+m_l^2}+7 m_X^2-14 m_X {E_\gamma}+6 {E^2_\gamma}\right)+2 m_l^2(5 m_X-2 {E_\gamma})\right) \right.\nonumber\\
		&~~+m_X^2(m_X-{E_\gamma})\left(-3 m_X^2\left(2 \sqrt{p^2_{l^-}+m_l^2}+{E_\gamma}\right)+2 m_X {E_\gamma}\left(6 \sqrt{p^2_{l^-}+m_l^2}+{E_\gamma}\right)-4 {E^2_\gamma} \sqrt{p^2_{l^-}+m_l^2}+m_X^3\right)\nonumber\\
		&~~+4 m_l^4\left(3 m_X^2-2 m_X {E_\gamma}+{E^2_\gamma}\right)+8 p^4_{l^-} m_X^2 \nonumber\\
		&~~+m_l^2 m_X \left.\left(-4 m_X^2\left(5 \sqrt{p^2_{l^-}+m_l^2}+8 {E_\gamma}\right)+2 m_X {E_\gamma}\left(14 \sqrt{p^2_{l^-}+m_l^2}+9 {E_\gamma}\right)-8 {E^2_\gamma} \sqrt{p^2_{l^-}+m_l^2}+15 m_X^3\right)\right) ~,
		\end{align}}where $\mathcal{Q}_{Q/\bar{Q}}$ and $\lambda_{Q/\bar{Q}}$ represent the electric charge and helicity of the $Q/\bar{Q}$ from the quarkonia $X$, respectively. 
Experimental findings show that for the vector particle $J/\Psi$, the spin projection along the $z$-axis is restricted to values of $\pm 1$, which corresponds to the state being represented as $|S,S_z \rangle=|1,\pm 1 \rangle$. This requirement ensures that $\lambda_Q=-\lambda_{\bar{Q}}$ at the leading order within the NRQCD framework.
Consequently, it is the squared amplitude where $\lambda_Q=-\lambda_{\bar{Q}}$ that exclusively contributes to the decay process of $J/\Psi$. By using the relation
	\begin{align}
	&\int d \Pi_f \left|\mathcal{M}(\lambda_Q=\lambda_{\bar{Q}};\vec{0}, \vec{0} \rightarrow \gamma l^+ l^-)\right|^2 = \frac{1}{2}  \int d \Pi_f \left|\mathcal{M}(\lambda_Q=-\lambda_{\bar{Q}};\vec{0}, \vec{0} \rightarrow \gamma l^+ l^-)\right|^2~,\label{rel}
	\end{align}
we get
	\begin{align}
	&\int d \Pi_f \left|\mathcal{M}(\lambda_Q=-\lambda_{\bar{Q}};\vec{0}, \vec{0} \rightarrow \gamma l^+ l^-)\right|^2=\frac{1}{3}  \int d \Pi_f \sum_{\lambda_Q,\lambda_{\bar{Q}}=\pm 1}\left|\mathcal{M}(\vec{0}, \vec{0} \rightarrow \gamma l^+ l^-)\right|^2~.
	\end{align}
	
\begin{figure}[htb]
		\begin{center}
			\includegraphics[width=0.49\textwidth]{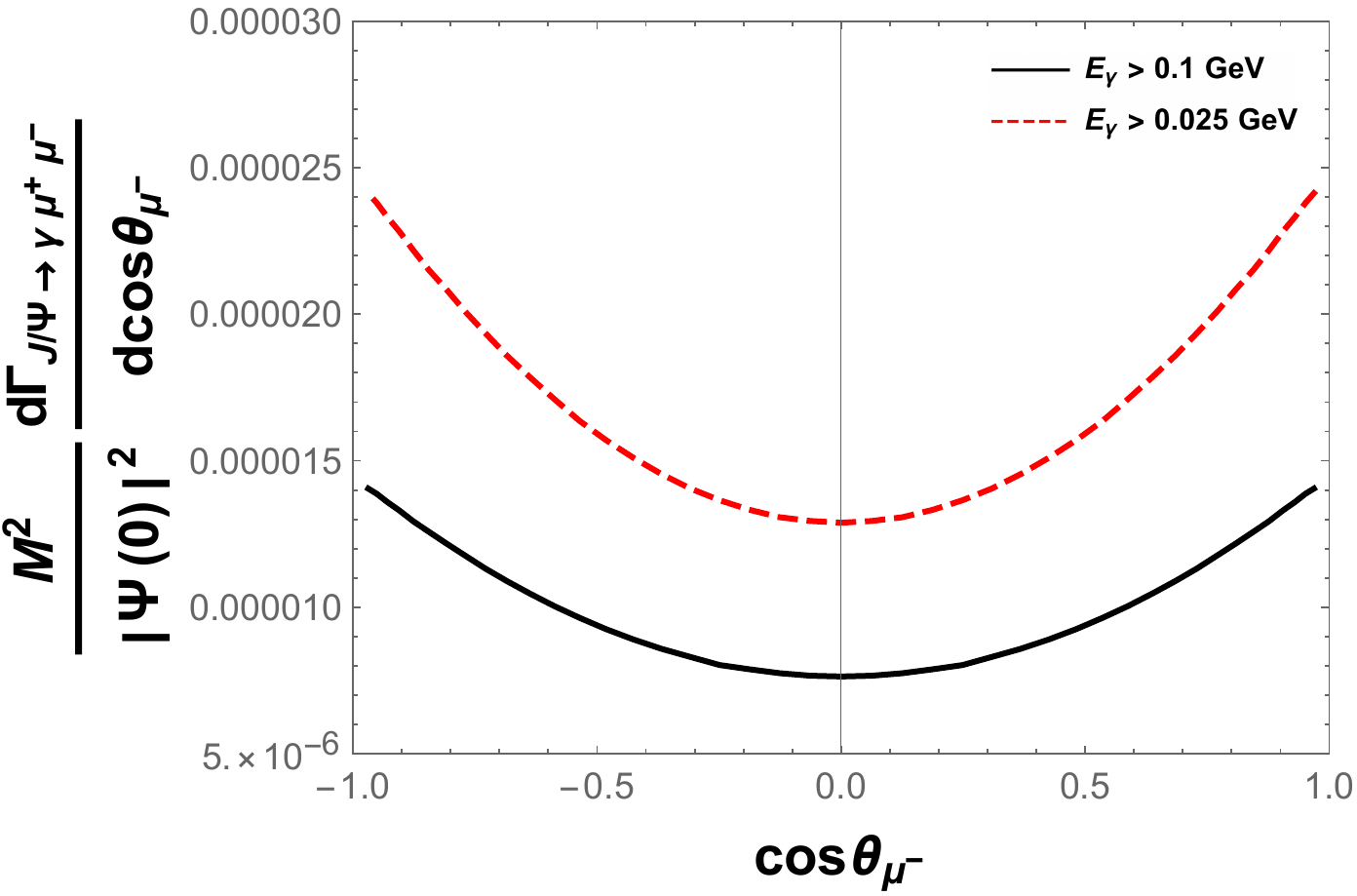}
			\includegraphics[width=0.49\textwidth]{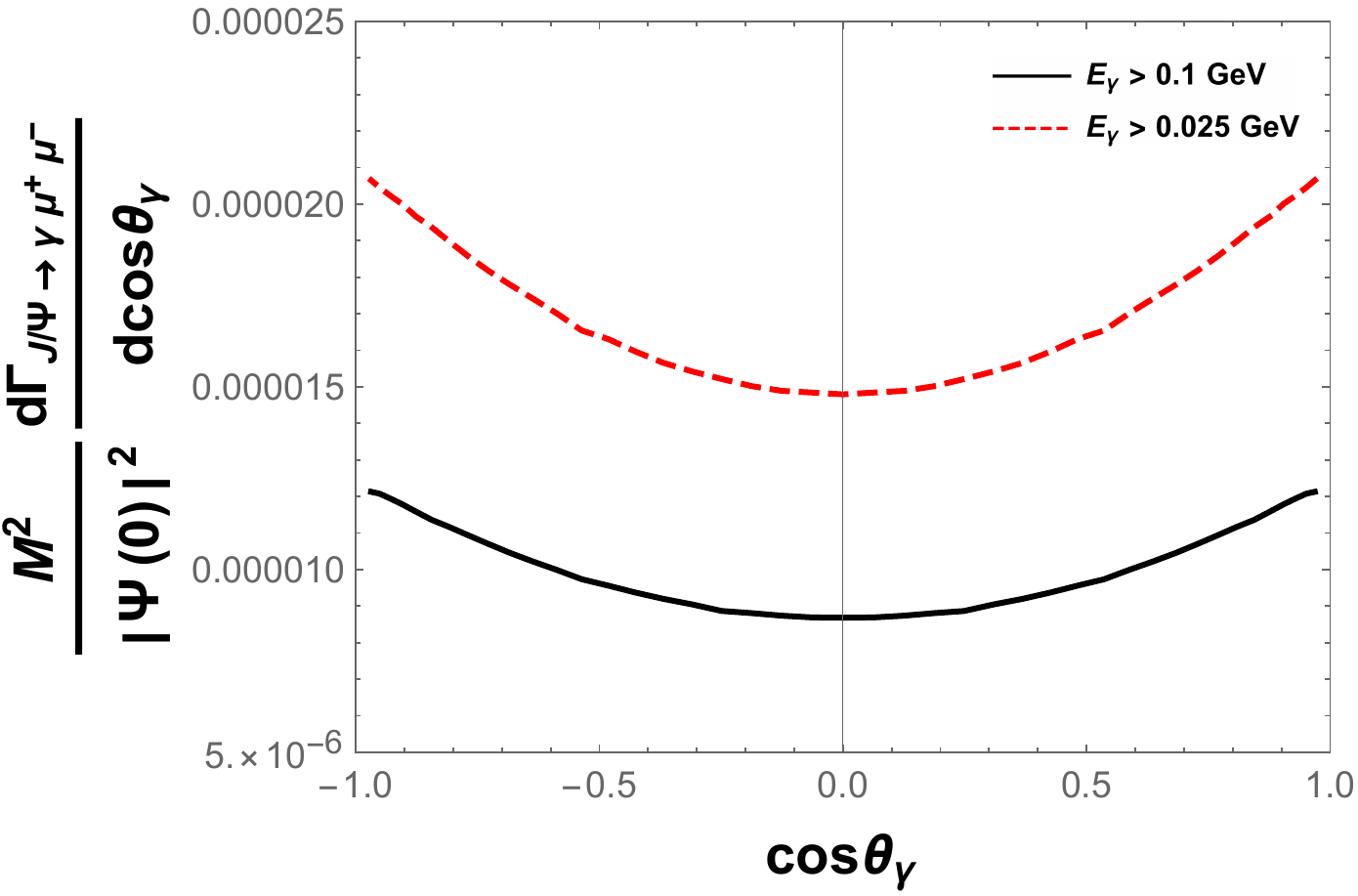}\\   
			\includegraphics[width=0.49\textwidth]{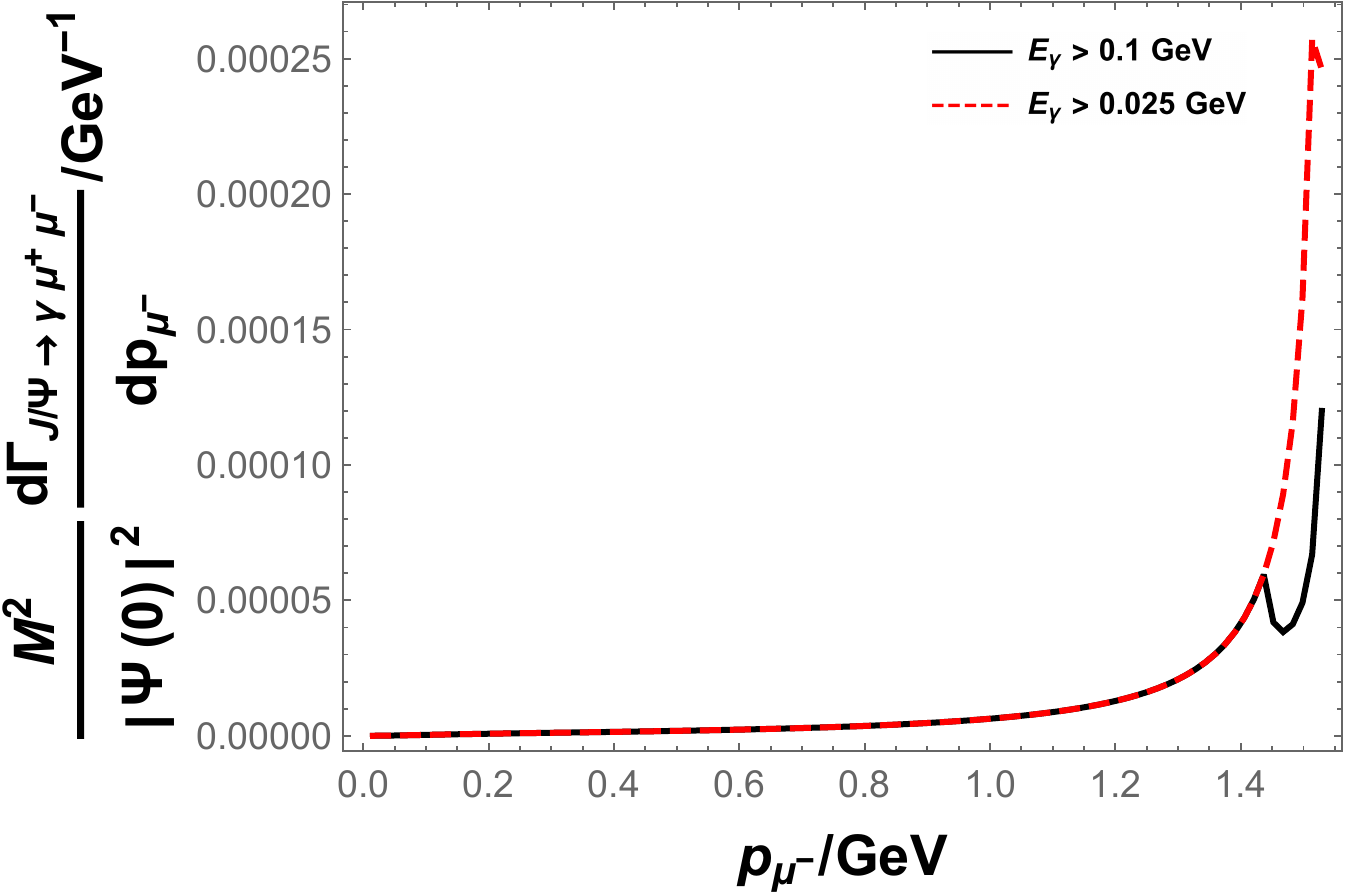}
			\includegraphics[width=0.49\textwidth]{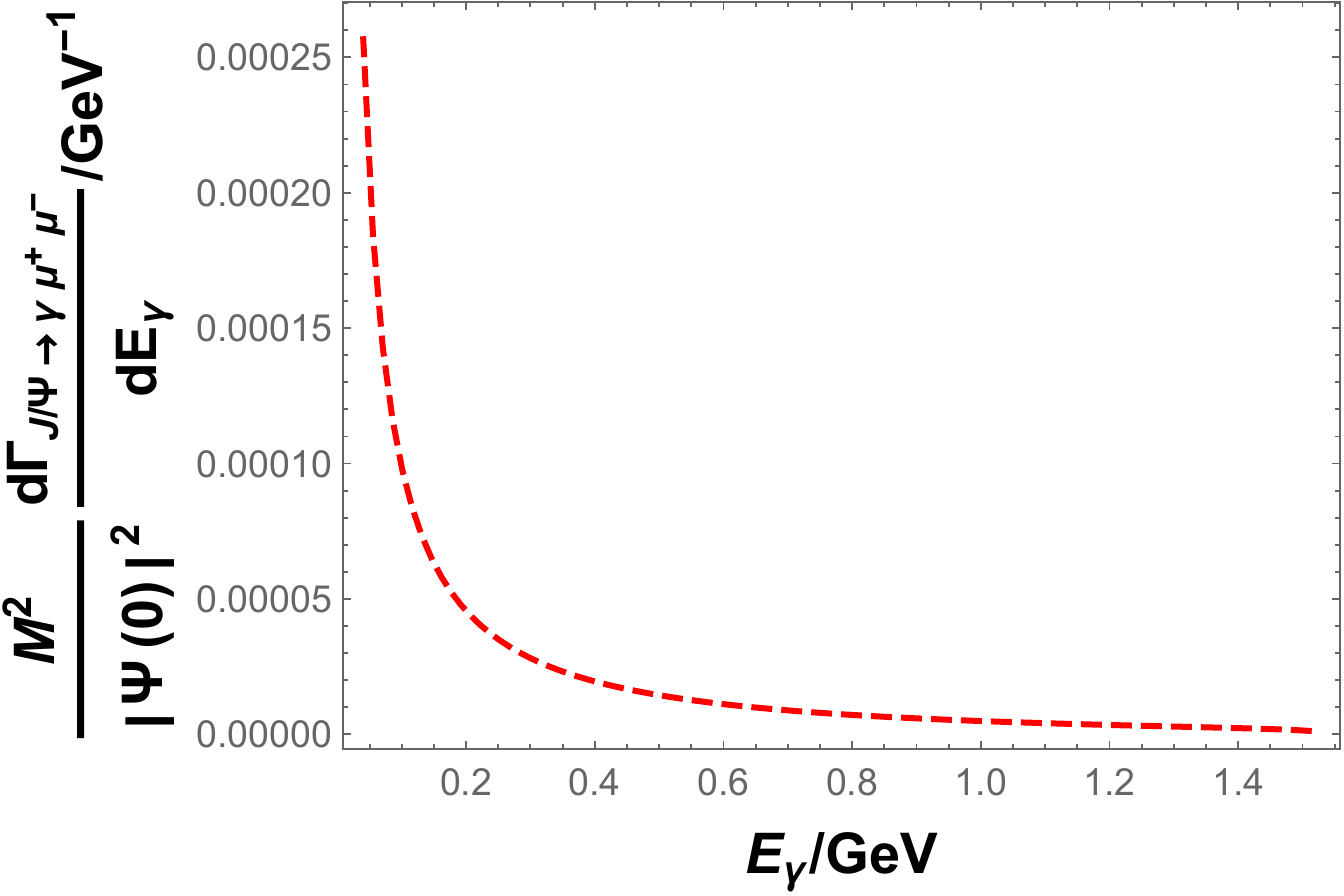}\\
			\includegraphics[width=0.49\textwidth]{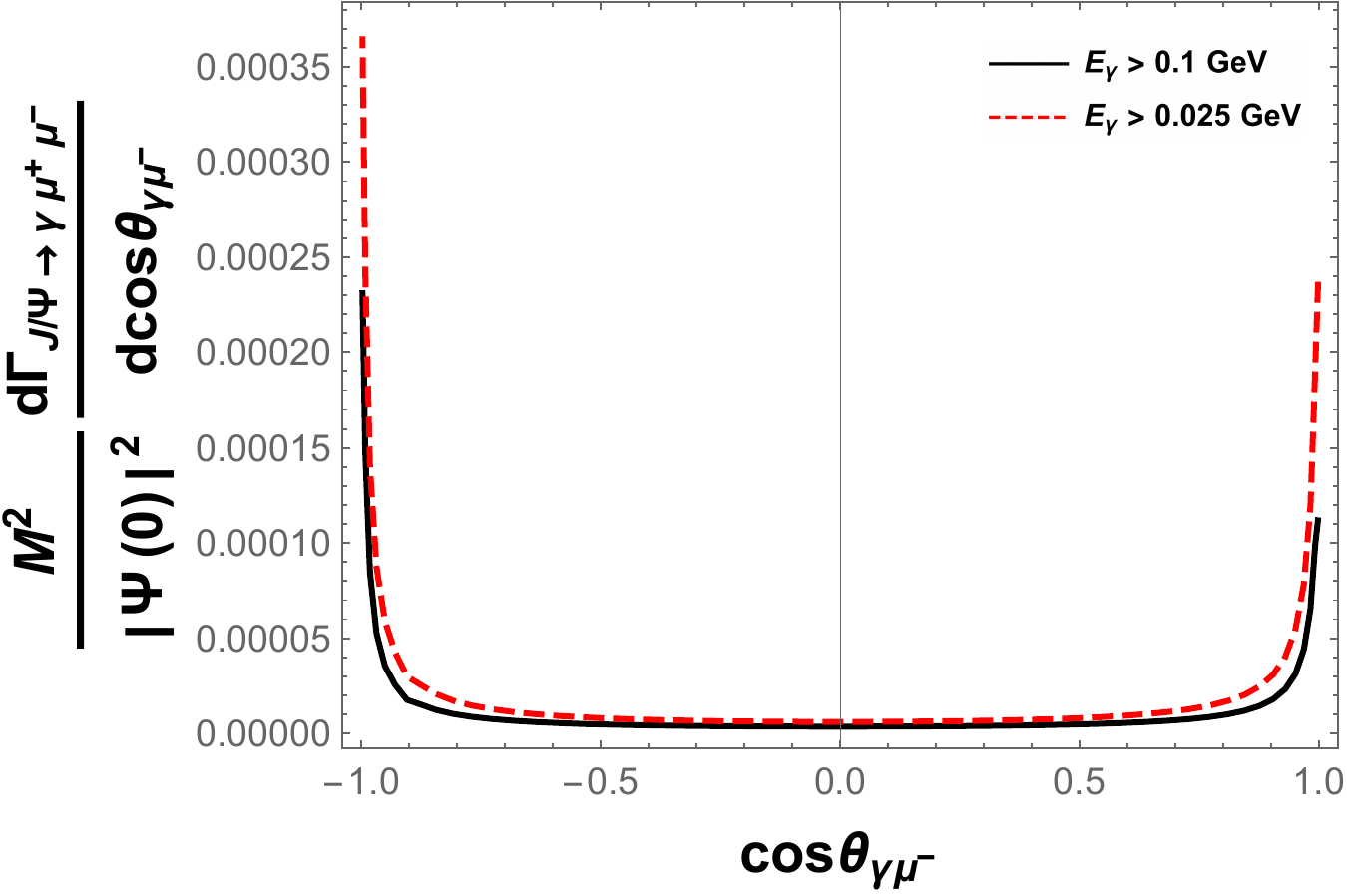}
		\end{center}
		\caption{\label{1dd} Angular and energy/momentum distributions of $\gamma$ and $\mu^-$ from the decay $J/\Psi \to \gamma \mu^+ \mu^-$ in the $J/\Psi$ rest frame. $\theta_{\gamma}$, $\theta_{\mu^-}$, and $\theta_{\gamma\mu^-}$ are polar angles of $\gamma$ and $\mu^-$, and the angle between the moving directions of $\gamma$ and $\mu^-$, respectively. $E_\gamma$ and $p_{\mu^-}$ are the photon energy and $\mu^-$ momentum, respectively. Two kinds of photon energy cut are chosen.}
	\end{figure}

\begin{figure}[htb]
		\begin{center}
			\includegraphics[width=0.49\textwidth]{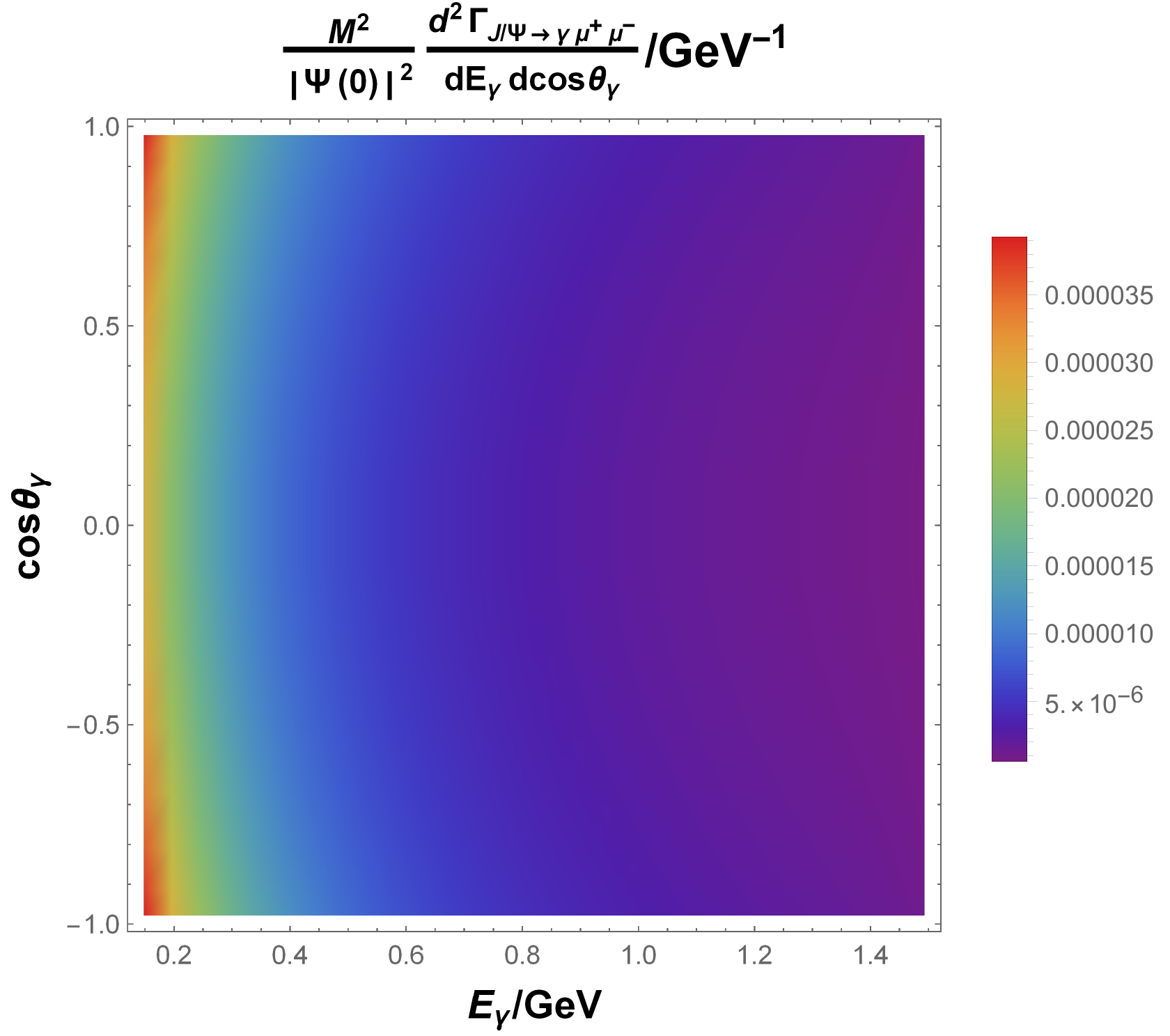}
			\includegraphics[width=0.49\textwidth]{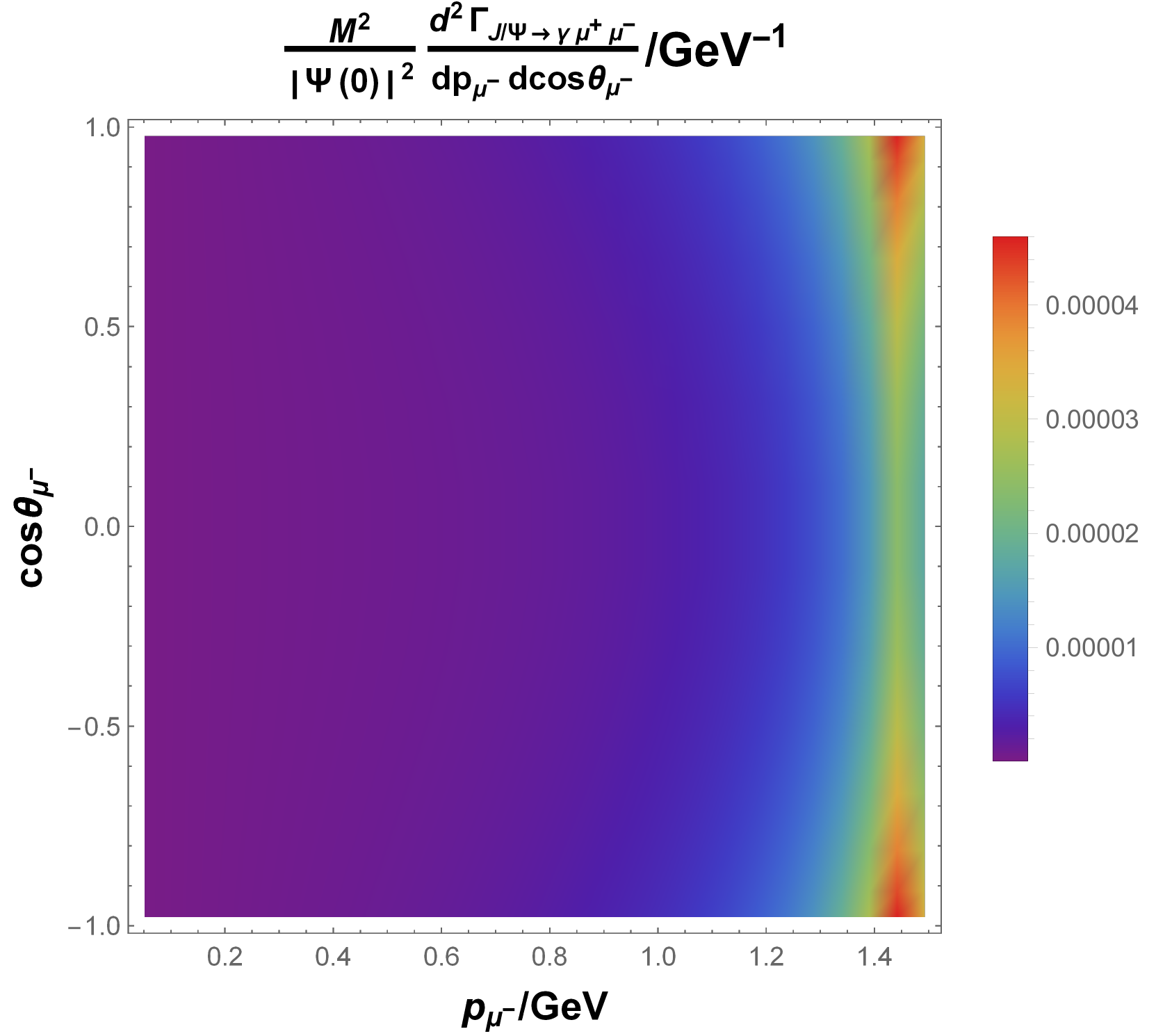}
		\end{center}
		\caption{\label{2dd} Left panel: polar angle-energy distribution of $\gamma$ from the decay $J/\Psi \to \gamma \mu^+ \mu^-$ in the $J/\Psi$ rest frame at $E_{\gamma \text{cut}}=0.1$ GeV. Right panel: polar angle-momentum distribution of $\mu^-$ from the decay $J/\Psi \to \gamma \mu^+ \mu^-$ in the $J/\Psi$ rest frame at $E_{\gamma \text{cut}}=0.1$ GeV.}
	\end{figure}

As an example, let us consider the decay $J/\Psi \to \gamma \mu^+ \mu^-$. By considering the squared amplitude with $\lambda_c=-\lambda_{\bar{c}}$ and the phase space in the rest frame of $J/\Psi$, numerical integration yields the angular and energy/momentum distributions of the final state particles, as shown in FIG.~\ref{1dd} with $E_{\gamma\text{cut}}=0.025,~0.1$ GeV and in FIG.~\ref{2dd} with $E_{\gamma\text{cut}}=0.1$ GeV.
{In the left middle panel of FIG. \ref{1dd}, noticeable kinks are observed in the lepton momentum ($p_{\mu^-}$) distributions. These kinks arise due to a shift in the permissible range of $E_\gamma$ values when $p_{\mu^-}$ surpasses $p_{l\text{mid}}$, as detailed in Eq. (\ref{errr}). The separation between the kink location ($p_{l\text{mid}}$) and the maximum value of $p_{\mu^-}$ ($p_{l\text{max}}$) is  $E_{\gamma\text{cut}}$, in accordance with Eq. (\ref{prange}). 
Regarding the photon energy distribution depicted in the right middle panel of FIG. \ref{1dd}, varying the photon energy cutoff ($E_{\gamma\text{cut}}$) merely alters the position of the left endpoint of the curve. It is noted that the distribution tends towards divergence as the photon energy approaches zero.
In the lowest panel of FIG.~\ref{1dd}, it is evident that radiative photons predominantly align along the collinear ($\theta_{\gamma\mu^-}= 0$) and antiparallel ($\theta_{\gamma\mu^-}= \pi$) directions relative to the $\mu^-$ movement, a phenomenon attributed to collinear singularity. Notably, the distribution exhibits asymmetry, with a higher propensity for photons to align in the antiparallel ($\theta_{\gamma\mu^-}= \pi$) direction to the $\mu^-$ trajectory. This asymmetry is due to the uneven phase space available at $\theta_{\gamma\mu^-}= 0$ and $\theta_{\gamma\mu^-}= \pi$. Specifically, with an $E_{\gamma\text{cut}}$ set at 0.1 GeV, the allowed ranges for $p_{\mu^-}$ are [0, 1.44] GeV and [0, 1.54] GeV at $\theta_{\gamma\mu^-}= 0$ and $\theta_{\gamma\mu^-}= \pi$, respectively.
}	
	
	\section{Expression of ${\Gamma_{X \rightarrow \gamma l^{+} l^{-}}}$ derived in the $l^+ l^-$ c.m. frame}\label{Section4}  
	
	Although the final states consist of three particles ($l^+$, $l^-$, and $\gamma$), it is advantageous to analyze the decay of $X$ in the c.m. frame of the $l^+ l^-$ system. In this frame, we can employ a set of variables to describe the angular distributions of the photon and leptons, as well as their relative motion. Specifically, we introduce $\Omega_\gamma^{\prime}$ ($\theta_\gamma^\prime$ and $\phi_\gamma^\prime$) and $\Omega_l^{\prime}$ ($\theta_l^\prime$ and $\phi_l^\prime$) to represent the angles of the photon and leptons, respectively, in the $l^+ l^-$ c.m. frame. Furthermore, we use $\theta^\prime_{\gamma l}$ to denote the angle between the directions of motions of the photon and the lepton in this frame. To quantify the kinematics, we introduce $\beta^\prime$, $E_\gamma^\prime$, and $s^\prime$ to represent the velocity of the leptons, the energy of the photon, and the square of the invariant mass of the $l^+ l^-$ system, respectively. By using the given quantities, we can parameterize the four-momentum of the final states in the c.m. frame of $l^+ l^-$ as follows:	
	\begin{align}
	&p_{l^-}^{\prime \mu} = (\sqrt{s^\prime}/{2},~0,~0,~\sqrt{{s^\prime}/{4}-m_l^2})~,\\
	&p_{l^+}^{\prime \mu} =(\sqrt{s^\prime}/{2},~0,~0,~-\sqrt{{s^\prime}/{4}-m_l^2})~,\\
	&p_{\gamma}^{\prime \mu} = (E_\gamma^\prime,~E_\gamma^\prime\sin\theta^\prime_{\gamma l}\cos\phi_\gamma^\prime,~E_\gamma^\prime\sin\theta^\prime_{\gamma l}\sin\phi_\gamma^\prime,~E_\gamma^\prime\cos\theta^\prime_{\gamma l})~,
	\end{align}
	where we have chosen the direction of motion of $l^-$ as the z-axis direction, resulting in $\theta_\gamma^\prime=\theta^\prime_{\gamma l}$.
	Using the relation $p^{\prime 2}_X=(p_{l^-}^{\prime}+p_{l^+}^{\prime}+p_{\gamma}^{\prime})^2=m_X^2$, we find
	\begin{align}
	&s^\prime=\left(\sqrt{m_X^2+E_\gamma^{\prime 2}}-E_\gamma^{\prime} \right)^2~, \\
	&\beta^\prime= \sqrt{1-\frac{4m_l^2}{\left(\sqrt{m_X^2+E_\gamma^{\prime 2}}-E_\gamma^{\prime} \right)^2}}~.
	\end{align}
	To ensure $s^\prime > 4m_l^2$, we obtain the upper limit for $E_\gamma^\prime$ as
	\begin{align}
	E_\gamma^\prime < \frac{m_X^2}{4m_l}-m_l~.
	\end{align}
	The photon energy cut $E_{\gamma\text{cut}}$ in the $X$ rest frame corresponds to the lower limit $E^\prime_{\gamma\text{cut}}$ for $E_\gamma^\prime$ in the $l^+ l^-$ c.m. frame, given by
	\begin{align}
	E^\prime_{\gamma\text{cut}}=E_{\gamma\text{cut}} \sqrt{\frac{1}{1-2\frac{E_{\gamma\text{cut}}}{m_X}}}<E_\gamma^\prime~.
	\end{align}
	The parameter of the Lorentz transformation from the $X$ rest frame to the $l^+ l^-$ c.m. frame is given by
	\begin{align}
	\vec{\beta}=-\frac{\vec{p}^\prime_\gamma}{\sqrt{m_X^2+E_\gamma^{\prime 2}}}~,
	\end{align} 
	where $\vec{p}^\prime_\gamma$ is the three-momentum of the photon in the $l^+ l^-$ c.m. frame. Thus, the $l^+ l^-$ c.m. frame changes with the photon energy and angles, and we need to determine the Jacobi factor $J$ appearing in the following equation:
	\begin{align}
	&\frac{d^3\vec{p}_{l^-}}{(2\pi)^3 2E_{l^-}}\frac{d^3\vec{p}_{l^+}}{(2\pi)^3 2E_{l^+}}\frac{d^3\vec{p}_{\gamma}}{(2\pi)^3 2E_{\gamma}}(2\pi)^4\delta^4(p^\mu_{X}-p^\mu_{l^-}-p^\mu_{l^+}-p^\mu_{\gamma})\nonumber\\
	=&J\times\frac{d^3\vec{p}^\prime_{l^-}}{(2\pi)^3 2E^\prime_{l^-}}\frac{d^3\vec{p}^\prime_{l^+}}{(2\pi)^3 2E^\prime_{l^+}}\frac{d^3\vec{p}^\prime_{\gamma}}{(2\pi)^3 2E^\prime_{\gamma}}(2\pi)^4\delta^4(p^{\prime\mu}_{X}-p^{\prime\mu}_{l^-}-p^{\prime\mu}_{l^+}-p^{\prime\mu}_{\gamma})~,
	\end{align}
	where $\vec{p}_{l^-}$, $\vec{p}_{l^+}$, and $\vec{p}_{\gamma}$ ($\vec{p}^\prime_{l^-}$, $\vec{p}^\prime_{l^+}$, and $\vec{p}^\prime_{\gamma}$) are the three-momenta of $l^-$, $l^+$, and $\gamma$ in the $X$ rest frame ($l^+ l^-$ c.m. frame), respectively. Direct derivations demonstrate that the Jacobi factor $J$ is related to the partial derivatives of the four-momentum of $l^-$ and $\gamma$ in the $X$ rest frame with respect to those in the $l^+ l^-$ c.m. frame, giving us
	\begin{align}	
	J=J(p^\mu_{l^-},p^\mu_{\gamma};p^{\prime\mu}_{l^-},p^{\prime\mu}_{\gamma})=J(E_\gamma^\prime)=
	1+\frac{E^{\prime 2}_\gamma}{m_X^2}\left(4-\frac{E^{\prime}_\gamma}{\sqrt{m_X^2+E^{\prime 2}_\gamma}}-3\frac{\sqrt{m_X^2+E^{\prime 2}_\gamma}}{E^{\prime}_\gamma} \right)~.
	\end{align}
	Therefore, we have
	\begin{align}
	&J(E_\gamma^\prime \ll m_X) =1~,\\
	&J(E_\gamma^\prime \gg m_X) =0~,
	\end{align}
	which indicates that the production of high-energy photons in the $l^+ l^-$ c.m. frame is suppressed.
	
	According to Eq. (\ref{decayx}), the decay width of quarkonia $X$ decaying into specific final states is directly proportional to the squared amplitude of the wave function at the origin, represented as $|\psi(\vec{0})|^2$. Consequently, the ratio between the decay widths of different decay modes of $X$ is independent of the specific value of $|\psi(\vec{0})|^2$. By taking into account the phase space, the Jacobi factor resulting from the Lorentz transformation, 
	the squared amplitude in the center-of-mass frame of the $l^+ l^-$ pair, and integrating over the final state angles, we derive the following expression:
	\begin{align}
	\frac{\Gamma_{X \rightarrow \gamma l^{+} l^{-}}}{\Gamma_{X \rightarrow l^{+} l^{-}}}=&{\frac{8\alpha b}{{\pi (1+2b)\sqrt{1-4b}}}} \int_{0}^{\sqrt{1-\frac{4b}{(\sqrt{1+c^2}-c)^2}}}d\beta^\prime \frac{
		\beta^\prime}{(1-\beta^{\prime 2})\left( 1-\beta^{\prime 2}-4b\right) } \nonumber\\
	&\times \left(
	-\beta^\prime{\left( 1+b^2\frac{16(2-\beta^{\prime 2})}{(1-\beta^{\prime 2})^2}\right) } +{\left((1-4b)+b^2\frac{8(3-\beta^{\prime 4})}{(1-\beta^{\prime 2})^2} \right)}\ln \left[ \frac{1+\beta^\prime}{1-\beta^\prime}\right]  
	\right)\equiv g^\prime(b,c)~, \label{3me} 
	\end{align}	
where 
\begin{align}
	&b=\frac{m_l^2}{m_X^2}~,\\
	&c=\frac{E_{\gamma\text{cut}}}{m_X}\sqrt{\frac{1}{1-2\frac{E_{\gamma\text{cut}}}{m_X}}}~.
\end{align} 
{The analytical solution for the integration outlined in Eq. (\ref{3me}) can be found in Appendix~{\ref{appen}}}.

	\begin{figure}[htb]
		\begin{center}
		\includegraphics[width=0.49\textwidth]{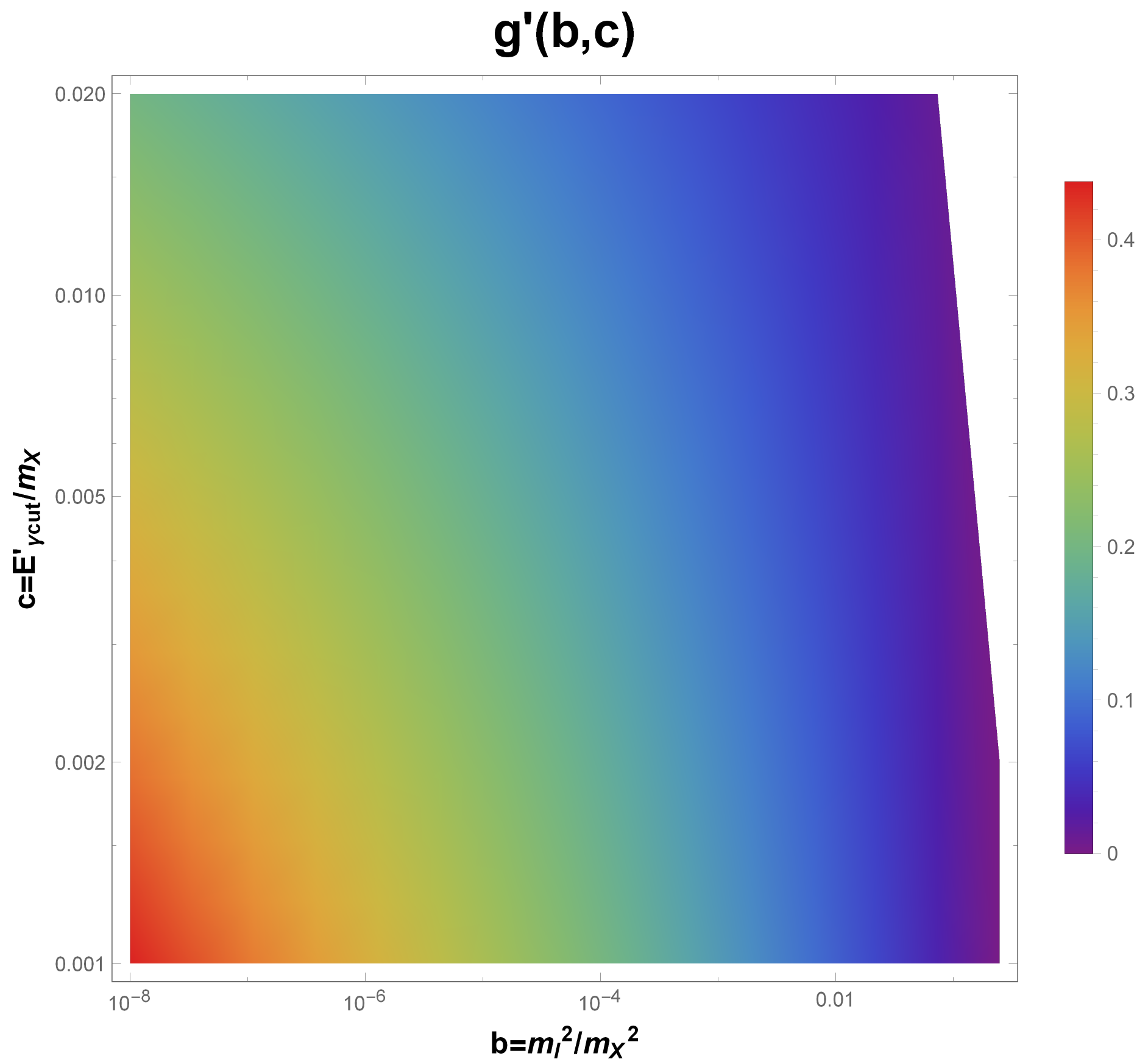}\\	\includegraphics[width=0.49\textwidth]{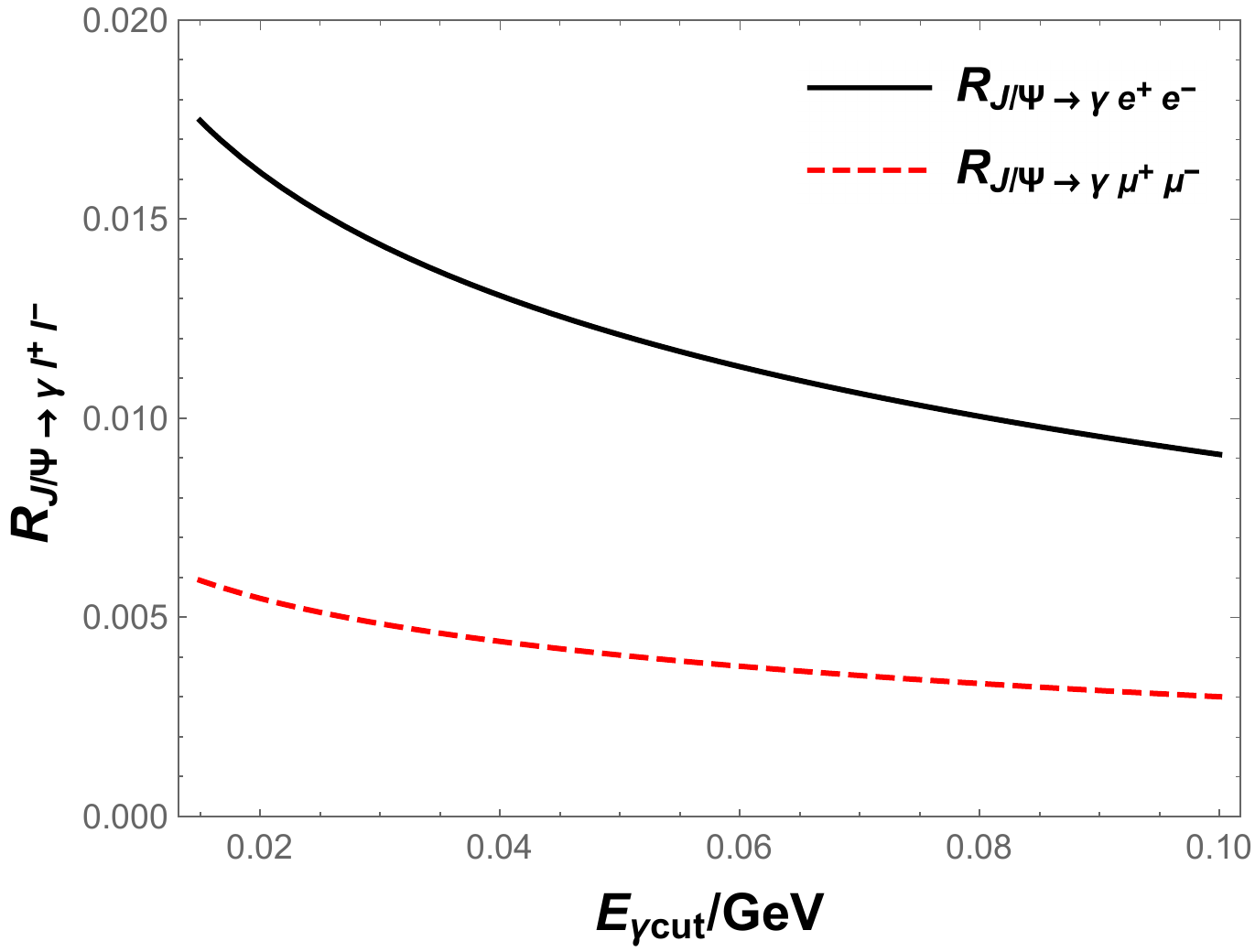}
			\includegraphics[width=0.49\textwidth]{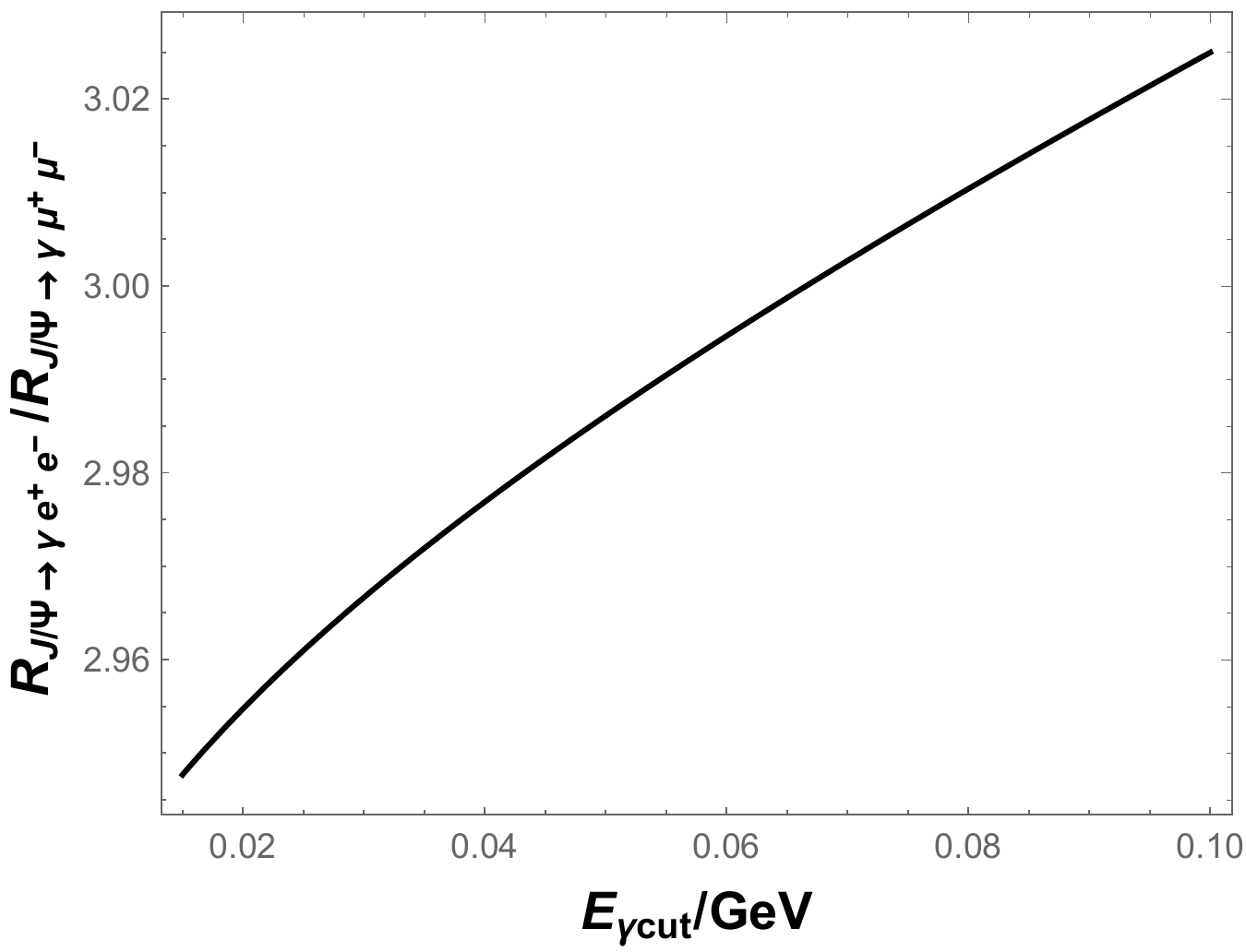}
		\end{center}
		\caption{\label{2cc} {Upper panel: the predicted values of ${\Gamma_{X \rightarrow \gamma l^{+} l^{-}}}/{\Gamma_{X \rightarrow l^{+} l^{-}}}$ using $g^\prime (b,c)$.} Lower left panel: branching ratios of $J/\Psi \to \gamma e^+ e^-$ and $J/\Psi \to \gamma \mu^+ \mu^-$ at different $E_{\gamma \text{cut}}$. Lower right panel: the ratio between $R_{J/\Psi \to \gamma e^+ e^-}$ and $R_{J/\Psi \to \gamma \mu^+ \mu^-}$ at different $E_{\gamma \text{cut}}$.}
	\end{figure}
	{The predicted ratios of ${\Gamma_{X \rightarrow \gamma l^{+} l^{-}}}/{\Gamma_{X \rightarrow l^{+} l^{-}}}$, calculated using $g^\prime (b,c)$, are depicted in FIG.~\ref{2cc}. These ratios exhibit a decrease as either the lepton mass $m_l$ or the photon energy cut $E_{\gamma\text{cut}}$ increases.}
	Utilizing the experimentally measured values of $R_{J/\Psi \to e^+ e^-}=(5.971 \pm 0.032) \%$ and $R_{J/\Psi \to \mu^+ \mu^-}=(5.961 \pm 0.033) \%$ obtained from the PDG \cite{ParticleDataGroup:2022pth}, and employing the ratio ${\Gamma_{X \rightarrow \gamma l^{+} l^{-}}}/{\Gamma_{X \rightarrow l^{+} l^{-}}}$ derived from Eq. (\ref{3me}), we can calculate the branching ratios of $J/\Psi \to \gamma e^+ e^-$ and $J/\Psi \to \gamma \mu^+ \mu^-$ at various $E_{\gamma \text{cut}}$ values, as depicted in FIG.~\ref{2cc}. Notably, it is noteworthy that the ratio between $R_{J/\Psi \to \gamma e^+ e^-}$ and $R_{J/\Psi \to \gamma \mu^+ \mu^-}$ consistently remains around 3.0 within the range of $E_{\gamma\text{cut}}$ spanning from $\mathcal{O}(0.01)$ to $\mathcal{O}(0.1)$ GeV. The future experimental results of this ratio will serve as a direct test of lepton flavor universality (LFU). {A more comprehensive analysis of this ratio, encompassing the robustness of theoretical predictions in the face of radiative and power corrections, the susceptibility to possible breaches of lepton flavor universality, and the practicality of experimentally detecting such BSM phenomena via this decay rate ratio, is beyond the scope of the current study. These facets are designated for subsequent research endeavors.}
	
	In Ref. \cite{FermilabE760:1996jsx}, the differential decay width for the process $X \rightarrow \gamma l^{+} l^{-}$ is given by
	\begin{align}
	&d \Gamma_{X \rightarrow \gamma l^{+} l^{-}}=  d \Gamma_{X \rightarrow l^{+} l^{-}} \beta^{\prime 3} \frac{2 \alpha}{\pi} \frac{d E_\gamma^{\prime}}{E_\gamma^{\prime}} \frac{s^{\prime}}{m_X^2} \frac{1-\cos ^2 \theta_{\gamma l}^{\prime}}{\left(1-\beta^{\prime 2} \cos ^2 \theta_{\gamma l}^{\prime}\right)^2} d \Omega_\gamma^{\prime}~, \label{3body}
	\end{align}
	where	
	\begin{align}	
	d \Gamma_{X \rightarrow l^{+} l^{-}}=\frac{3}{3+\lambda}\left(1+\lambda \cos ^2 \theta_l^{\prime}\right) \Gamma_{X \rightarrow l^{+} l^{-}} \frac{d \Omega_l^{\prime}}{4 \pi}\label{2body}~.
	\end{align}
	Upon integrating out the final state angles ($\Omega^\prime_\gamma$ and $\Omega^\prime_l$), we obtain the ratio of the decay widths as
	\begin{align}
	\frac{\Gamma_{X \rightarrow \gamma l^{+} l^{-}}}{\Gamma_{X \rightarrow l^{+} l^{-}}}=&16\alpha b\int_{0}^{\sqrt{1-\frac{4b}{(\sqrt{1+c^2}-c)^2}}}d\beta^\prime {\frac{\beta^\prime}{(1-\beta^{\prime 2})\left( 1-\beta^{\prime 2}-4b\right) }}  \nonumber\\
	&\times{\left( -\beta^\prime{ \left( 1+b \frac{4}{1-\beta^{\prime 2}}\right)}+{\left( 1+b \frac{4}{1-\beta^{\prime 2}}\right)\frac{1+\beta^{\prime 2}}{2}}\ln \left[ \frac{1+\beta^\prime}{1-\beta^\prime} \right] \right)}\equiv g(b,c)
	~.\label{gjw}
	\end{align}	
The analytical result for the integration in Eq.~(\ref{gjw}) is provided in Appendix~{\ref{appen}}.

\begin{figure}[ht]
	\centering
	\includegraphics[width=0.8\textwidth]{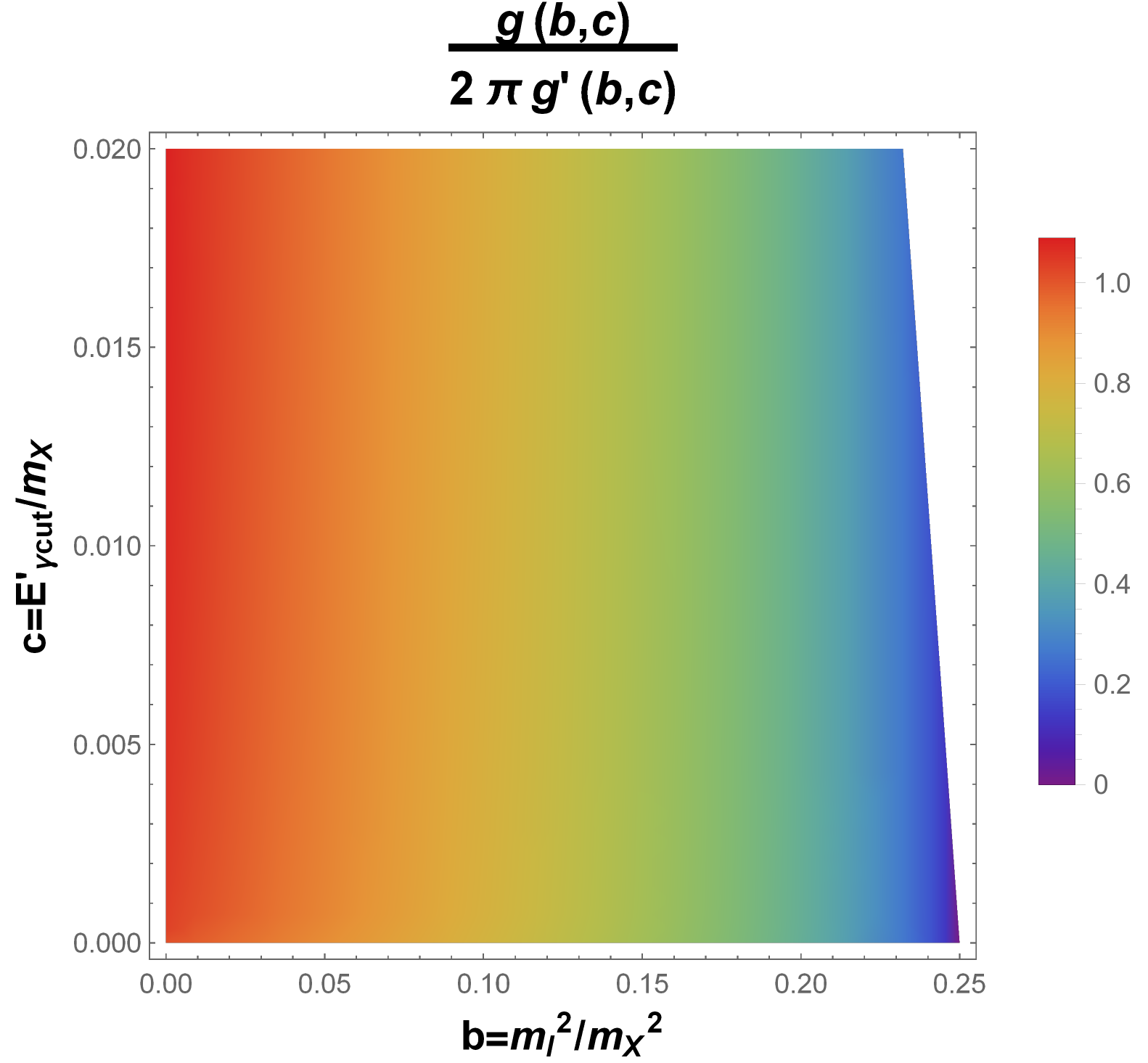}
	\caption{The ratio between the predicted values of ${\Gamma_{X \rightarrow \gamma l^{+} l^{-}}}/{\Gamma_{X \rightarrow l^{+} l^{-}}}$ using $\frac{g(c,b)}{2\pi}$ and $g^\prime (b,c)$.}
	\label{ggp}
\end{figure}

The comparison between $g(b,c)$ derived from the relation proposed in Ref. \cite{FermilabE760:1996jsx} and $g^\prime(b,c)$ obtained in this study reveals differences arising from the prefactors and integrated functions of $\beta^\prime$.

The dependencies of $\frac{g(b,c)}{2\pi g^\prime(b,c)}$ on $b$ and $c$ are illustrated in FIG.~\ref{ggp}. It is observed that the ratio $\frac{g(b,c)}{2\pi g^\prime(b,c)}$ is greater than one for small values of $b$ (e.g., $b<0.03$ when $c=0.02$) and decreases as $b$ increases.

In the specific case where $E_{\gamma \text{cut}}=0.1$ GeV and $X$ includes $J/\Psi$, $\Psi(2S)$, $\Upsilon(1S)$, and $\Upsilon(2S)$, the ratio between ${\Gamma_{X \rightarrow \gamma l^{+} l^{-}}}$ and ${\Gamma_{X \rightarrow l^{+} l^{-}}}$, as determined by $g(b,c)$ and $g^\prime (b,c)$, is presented in TABLE.~\ref{Benchmark points}. It is found that for modes with $b$ approximately equal to or smaller than $\mathcal{O}(10^{-2})$, $\frac{g(c,b)}{2\pi}$, rather than $g(c,b)$, is comparable to $g^\prime (c,b)$, with a relative difference within 10\%, which is comparable to the NLO corrections of $O(\alpha)$ and $O(v^4)$ according to Eq.~(\ref{nlo}). The largest discrepancy between $\frac{g(c,b)}{2\pi}$ and $g^\prime (c,b)$ is observed for the mode ${{\Psi(2S) \rightarrow (\gamma) \tau^{+} \tau^{-}}}$, where $b=0.232$, and $g^\prime (c,b)$ is four times larger than $\frac{g(c,b)}{2\pi}$. Compared to $\frac{g(c,b)}{2\pi}$, the factor $\frac{1}{(1+2b)\sqrt{1-4b}}$ in $g^\prime (c,b)$ yields a 2.5-fold enhancement for the decay mode ${{\Psi(2S) \rightarrow (\gamma) \tau^{+} \tau^{-}}}$ when $b=0.232$. By disregarding the differences in the prefactors between $\frac{g(c,b)}{2\pi}$ and $g^\prime (c,b)$, which are negligible when $b\ll 1$, the remaining enhancement in $g^\prime (c,b)$ for the ${{\Psi(2S) \rightarrow (\gamma) \tau^{+} \tau^{-}}}$ decay mode can be attributed to the variations in the integrated functions. 

\begin{table}[h]
	\begin{center}
		\begin{tabular}{ccccc}
			\hline
			Mode & $b$ & $c$ ($E_{\gamma\text{cut}}=0.1$ GeV) & $g^\prime(c,b)$ & $\frac{g(c,b)}{2\pi g^\prime(c,b)}$\\
			\hline
			${{J/\Psi \rightarrow (\gamma) e^{+} e^{-}}}$ & $2.72\times 10^{-8}$ & $3.34\times 10^{-2}$ & 0.152 & 1.11\\
			${{J/\Psi \rightarrow (\gamma) \mu^{+} \mu^{-}}}$ & $1.16\times 10^{-3}$ & $3.34\times 10^{-2}$ & $5.04\times 10^{-2}$  & 1.10\\
			${{\Psi(2S) \rightarrow (\gamma) e^{+} e^{-}}}$ & $1.92\times 10^{-8}$ & $2.79\times 10^{-2}$ & 0.168 & 1.10\\
			${{\Psi(2S) \rightarrow (\gamma) \mu^{+} \mu^{-}}}$ & $8.22\times 10^{-4}$ & $2.79\times 10^{-2}$ & $5.84\times 10^{-2}$ & 1.10\\
			${{\Psi(2S) \rightarrow (\gamma) \tau^{+} \tau^{-}}}$ & {0.232} & $2.79\times 10^{-2}$ & $6.09\times 10^{-6}$ & {0.25}\\
			${{\Upsilon(1S) \rightarrow (\gamma) e^{+} e^{-}}}$ & $2.92\times 10^{-9}$ & $1.07\times 10^{-2}$ & 0.266 & 1.08\\
			${{\Upsilon(1S) \rightarrow (\gamma) \mu^{+} \mu^{-}}}$ & $1.25\times 10^{-4}$ & $1.07\times 10^{-2}$ & 0.111 & 1.07\\
			${{\Upsilon(1S) \rightarrow (\gamma) \tau^{+} \tau^{-}}}$ & $3.53\times 10^{-2}$ & $1.07\times 10^{-2}$ & $2.80\times 10^{-2}$ & 0.98\\
			${{\Upsilon(2S) \rightarrow (\gamma) e^{+} e^{-}}}$ & $2.60\times 10^{-9}$ & $1.01\times 10^{-2}$ & 0.273 & 1.07\\
			${{\Upsilon(2S) \rightarrow (\gamma) \mu^{+} \mu^{-}}}$ & $1.11\times 10^{-4}$ & $1.01\times 10^{-2}$ & 0.115 & 1.07\\
			${{\Upsilon(2S) \rightarrow (\gamma) \tau^{+} \tau^{-}}}$ & $3.14\times 10^{-2}$ & $1.01\times 10^{-2}$ & $3.04\times 10^{-2}$ & 0.99\\
			\hline
		\end{tabular}
	\end{center}
	\caption{${\Gamma_{X \rightarrow \gamma l^{+} l^{-}}}/{\Gamma_{X \rightarrow l^{+} l^{-}}}$ determined by $g^\prime (b,c)$ and the ratio between the predicted values of ${\Gamma_{X \rightarrow \gamma l^{+} l^{-}}}/{\Gamma_{X \rightarrow l^{+} l^{-}}}$ using $\frac{g(c,b)}{2\pi}$ and $g^\prime (b,c)$. $X$ represents $J/\Psi$, $\Psi(2S)$, $\Upsilon(1S)$, and $\Upsilon(2S)$.
}
	\label{Benchmark points}
\end{table}

\section{Conclusion}\label{Section6}

{In this research, we delve into the leptonic decays of heavy quarkonia $X$, conceptualized as a zero-momentum bound state of a quark ($Q$) and an antiquark ($\bar{Q}$) within the NRQCD framework at the LO level. Through a detailed analysis of the squared amplitude for the decay process $X \to \gamma l^+ l^-$ and an examination of the kinematics of the final state particles in the rest frame of $X$, we have calculated the decay widths and branching ratios for the three-body leptonic decays of $X$. Our study also extends to determining the angular and energy/momentum distributions of the final state particles in the rest frame of $X$ via numerical integration.

To explore the Lorentz transformation from the rest frame of $X$ to the c.m. frame of $l^+ l^-$, we derived the Jacobian of this transformation and analyzed the phase space in the $l^+ l^-$ c.m. frame. These analytical efforts have led us to propose a prediction for the ratio between ${\Gamma_{X \rightarrow \gamma l^{+} l^{-}}}$ and ${\Gamma_{X \rightarrow l^{+} l^{-}}}$, highlighting discrepancies with previously reported expressions, notably from Ref. \cite{FermilabE760:1996jsx}. Our analysis identifies a missing factor of $1/2\pi$ in the earlier expression and suggests that a detailed comparison of the equations can elucidate the differences. Significantly, the discrepancies become pronounced when the ratio $m_l^2/m_X^2$ crosses a threshold of approximately $\mathcal{O}(0.1)$.

Our investigation spans various heavy quarkonia, including $J/\Psi$, $\Psi(2S)$, $\Upsilon(1S)$, and $\Upsilon(2S)$. The ratios derived from our proposed relation align closely with those from the literature (after adjusting by $2\pi$), showcasing relative differences generally within or below 10$\%$. This level of agreement is on par with the expected Next-to-Leading Order (NLO) corrections of $O(\alpha)$ and $O(v^4)$. An outlier is observed in the ratio between ${\Gamma_{\Psi(2S) \to \gamma \tau^+ \tau^-}}$ and ${\Gamma_{\Psi(2S) \to \tau^+ \tau^-}}$, where our proposed expression predicts a value four times larger than expected. The findings from this research offer a solid foundation for comparison with future experimental results from facilities like BESIII, B-factory experiments, CEPC, and others, thereby enhancing our understanding of heavy quarkonia decays and contributing to tests of the SM and the search for new physics.}

\appendix
\section{Expressions of $g^\prime(b,c)$ and $g(b,c)$}\label{appen}
The analytical solutions for the integrations described in Eqs. (\ref{3me}) and (\ref{gjw}) are 
{\footnotesize
\begin{align}
   g^\prime(b,c)=&\frac{\alpha }{\pi  (1+2b) \sqrt{1-4 b}}\times\left(
   \left(1-4 b^2\right) \left(\frac{\pi ^2}{3}-2 \ln ^2(2)\right)+\frac{2 b x \left(4+23b-(4+17 b) x^2\right)}{\left(1-x^2\right)^2}
   \right.\nonumber\\
   &+\frac{2 \left(2-23 b^2-2 (2+b (2-7 b)) x^2+(2+b (4+b)) x^4\right) \tanh ^{-1}(x)}{\left(1-x^2\right)^2}\nonumber\\
   &-\frac{4 \left(1-2 b-8 b^2\right) \tanh ^{-1}\left(\frac{x}{\sqrt{1-4 b}}\right)}{\sqrt{1-4 b}}\nonumber\\
   &+\left(1-4 b^2\right) \left(2 \ln \left(\frac{1-x}{1+x}\right) \ln \left(\frac{1-4 b-x^2}{4 b}\right)-\ln ^2\left(1-x^2\right)+2 \ln ^2(1+x)+4 \ln (2) \ln (1-x)\right)\nonumber\\
   &-2 \left(1-4 b^2\right) \left(2 \text{Li}_2\left(\frac{1-x}{2}\right)-\Re\left(\text{Li}_2\left(\frac{1-x}{1-\sqrt{1-4 b}}\right)\right)+\Re\left(\text{Li}_2\left(\frac{1+x}{1-\sqrt{1-4 b}}\right)\right)\right.\nonumber\\
   &-\left.\left.\text{Li}_2\left(\frac{1-x}{1+\sqrt{1-4 b}}\right)+\text{Li}_2\left(\frac{1+x}{1+\sqrt{1-4 b}}\right)\right)\right)
\end{align}}
and
{\footnotesize
\begin{align}
    g(b,c)=&2\alpha\times\left((1-b) \left(\frac{\pi ^2}{3}-2 \ln ^2(2)\right)+\frac{8 b x}{1-x^2}+4 b \tanh ^{-1}(x)-4 \sqrt{1-4 b} \tanh ^{-1}\left(\frac{x}{\sqrt{1-4 b}}\right) \right.\nonumber\\
    & +(1-b) \left(\frac{2 (1-2 b) \ln \left(\frac{1-\sqrt{1-4 b}}{1+\sqrt{1-4 b}}\right) \ln \left(\frac{\sqrt{1-4 b}-x}{\sqrt{1-4 b}+x}\right)}{1-b}-\ln ^2\left(1-x^2\right)+2 \ln ^2(1+x)+4 \ln (2) \ln (1-x)\right.\nonumber\\
    &\left.-\frac{2 \left(1-3b- \left(1-b\right)x^2\right) \ln \left(\frac{1-x}{1+x}\right)}{(1-b) \left(1-x^2\right)}\right)-4 (1-b) \text{Li}_2\left(\frac{1-x}{2}\right)\nonumber\\
    &-2 (1-2 b) \left.\left(\text{Li}_2\left(-\frac{\sqrt{1-4 b}-x}{1-\sqrt{1-4 b}}\right)-\text{Li}_2\left(-\frac{\sqrt{1-4 b}+x}{1-\sqrt{1-4 b}}\right)-\text{Li}_2\left(\frac{\sqrt{1-4 b}-x}{1+\sqrt{1-4 b}}\right)+\text{Li}_2\left(\frac{\sqrt{1-4 b}+x}{1+\sqrt{1-4 b}}\right)\right)\right)~,
\end{align}}
respectively, where
\begin{align}
    x=\sqrt{1-\frac{4b}{(\sqrt{1+c^2}-c)^2}}~,
\end{align}
and $\Re(z)$ represent the real part of $z$.
\newpage
\section*{Acknowledgement}
This work was supported by the National Natural Science Foundation of China under grants No.12247119 and No. 12042507.

\phantomsection
\addcontentsline{toc}{section}{References}

\end{document}